\documentclass[prl,twocolumn,showpacs,preprintnumbers,amsmath,amssymb]{revtex4}

\usepackage{graphicx}
\usepackage{dcolumn}
\usepackage{bm}

\begin{document}


\title{Superconductivity at 41 K and its competition with spin-density-wave instability in layered CeO$_{1-x}$F$_x$FeAs}
\author{G. F. Chen}
\author{Z. Li}
\author{D. Wu}
\author{G. Li}
\author{W. Z. Hu}
\author{J. Dong}
\author{P. Zheng}
\author{J. L. Luo}
\author{N. L. Wang}

\affiliation{Beijing National Laboratory for Condensed Matter
Physics, Institute of Physics, Chinese Academy of Sciences,
Beijing 100080, People¡¯s Republic of China}


\begin{abstract}

A series of layered CeO$_{1-x}$F$_x$FeAs compounds with x=0 to
0.20 are synthesized by solid state reaction method. Similar to
the LaOFeAs, the pure CeOFeAs shows a strong resistivity anomaly
near 145 K, which was ascribed to the spin-density-wave
instability. F-doping suppresses this instability and leads to the
superconducting ground state. Most surprisingly, the
superconducting transition temperature could reach as high as 41
K. The very high superconducting transition temperature strongly
challenges the classic BCS theory based on the electron-phonon
interaction. The very closeness of the superconducting phase to
the spin-density-wave instability suggests that the magnetic
fluctuations play a key role in the superconducting paring
mechanism. The study also reveals that the Ce 4f electrons form
local moments and ordered antiferromagnetically below 4 K, which
could coexist with superconductivity.
\end{abstract}



\maketitle

The recent discovery of superconductivity with transition
temperature of 26 K in LaO$_{1-x}$F$_x$FeAs
system\cite{Kamihara08} has generated tremendous interest in the
scientific community. Except for a relatively high transition
temperature, the system displays many interesting properties.
Among others, the presence of competing ordered ground states is
one of the most interesting phenomena\cite{Dong}. The pure LaOFeAs
itself is not superconducting but shows an anomaly near 150 K in
both resistivity and dc magnetic susceptibility.\cite{Kamihara08}
This anomaly was shown to be caused by the spin-density-wave (SDW)
instability.\cite{Dong} Electron-doping by F suppresses the SDW
instability and recovers the superconductivity. Here we show that
similar competing orders exist in another rear-earth transition
metal oxypnictide Ce(O$_{1-x}$F$_x$)FeAs. Most surprisingly, the
superconducting transition temperature in this system could reach
as high as 41 K. Except for cuprate superconductors, T$_c$ in such
iron-based compounds has already become the highest.

The very high superconducting transition temperature has several
important implications. First, the T$_c$ value has already reached
the well-accepted limit value of classic BCS
theory\cite{McMillan,Ginzburg}. Considering the small carrier
density and rather week electron-phonon coupling estimated from
first-principle calculations\cite{Boeri,Mazin}, the observation
result strongly challenges the BCS theory based on the
electron-phonon interaction. Second, the rare-earth Ce-based
compounds usually show hybridization between localized f-electrons
and itinerant electrons. This often leads to a strong enhancement
of carrier effective mass at low temperature. Even for 4d
transition metal oxypnictide with the same type of structure, a
recent report indicates that the electronic specific heat
coefficient of Ce-based CeORuP ($\gamma$=77 mJ/mol K$^2$) is 20
times higher than the value of La-based LaORuP ($\gamma$=3.9
mJ/mol K$^2$)\cite{Krellner}. The hybridization also tends to
cause various ordered states at low temperature, like
ferromagnetic (FM) or antiferromagnetic (AFM) ordering. Although
superconducting state could occur in Ce-based materials, the
superconducting transition temperature is usually very low. The
highest superconducting transition temperature is only 2.3 K
achieved in CeCoIn$_5$\cite{Petrovic}. The extremely high
superconducting transition temperature obtained here on
Ce(O$_{1-x}$F$_x$)FeAs offers an opportunity to examine the role
played by Ce 4f electrons. Third, the superconducting phase is
very close to the spin-density-wave instability. This indicates
that the magnetic fluctuations plays a key role in the
superconducting pairing mechanism.

\begin{figure}[b]
\includegraphics[width=8cm,clip]{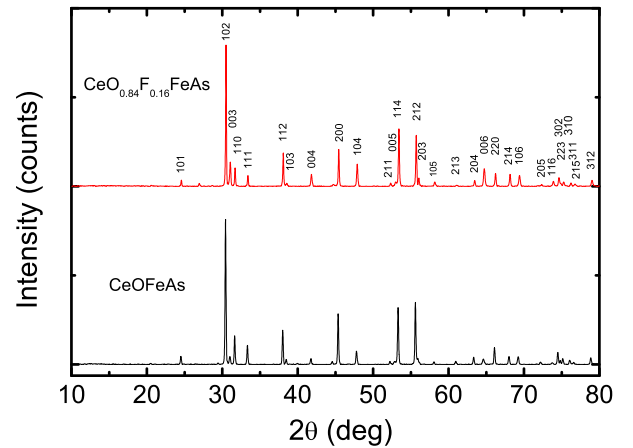}
\caption{(Color online) X-ray powder diffraction patterns of the
pure CeOFeAs and CeO$_{0.84}$F$_{0.16}$FeAs compounds.}
\end{figure}

A series of layered CeO$_{1-x}$F$_x$FeAs compounds with x=0, 0.04,
0.08, 0.12, 0.16 and 0.20 are synthesized by solid state reaction
method using CeAs, Fe, CeO$_2$, CeF$_3$, Fe$_2$As as starting
materials. CeAs was obtained by reacting Ce chips and As pieces at
500 $^{\circ}C$ for 15 hours and then 850 $^{\circ}C$ for 5 hours.
The raw materials were thoroughly mixed and pressed into pellets.
The pellets were wrapped with Ta foil and sealed in an evacuate
quartz tube. They were then annealed at 1150 $^{\circ}C$ for 50
hours. The resulting samples were characterized by a powder X-ray
diffraction(XRD) method with Cu K$\alpha$ radiation at room
temperature. Almost pure phase was achieved for those samples. The
XRD patterns for the parent x=0 and x=0.16 compounds are shown in
Fig. 1, which could be well indexed on the basis of tetragonal
ZrCuSiAs-type structure with the space group P4/nmm. The lattice
parameters for the parent and x=0.16 compounds obtained by a
least-square fit to the experimental data are a=3.996 $\AA$,
c=8.648 $\AA$, and a=3.989 $\AA$, c=8.631 $\AA$, respectively.
Compared to the undoped phase CeOFeAs, the apparent reduction of
the lattice volume upon F-doping indicates a successful chemical
substitution.

\begin{figure}[t]
\centerline{\includegraphics[width=3.2in]{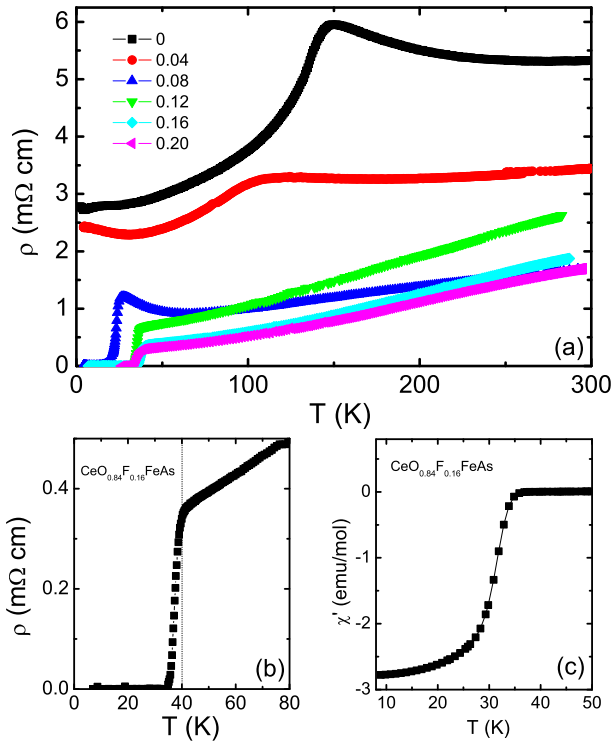}}%
\caption{(Color online) (a) The electrical resistivity vs
temperature for a series of CeO$_{1-x}$F$_{x}$FeAs. (b)
T-dependent resistivity in an expanded region for x=0.16 sample.
The superconducting transition with sharp onset temperature at 41
K is seen. (c) Real part of T-dependent ac magnetic
susceptibility.}
\end{figure}

Standard 4-probe dc resistivity and ac susceptibility measurements
were preformed down to 1.8K in a Physical Property Measurement
System(PPMS) of Quantum Design company. Figure 2 (a) shows the
temperature dependence of the resistivity. The pure CeOFeAs sample
has rather high dc resistivity value. The resistivity increases
slightly with decreasing temperature, but below roughly 145 K, the
resistivity drops steeply. After F-doping, the overall resistivity
decreases and the 145 K anomaly shifts to the lower temperature
and becomes less pronounced. At higher F-doping, the anomaly
disappears and a superconducting transition occurs. The highest
T$_c$=41 K is obtained at x=0.16, which can be seen clearly from
an expanded plot of the temperature-dependent resistivity curve.
T$_c$ drops slightly with further F-doping. The bulk
superconductivity in F-doped CeOFeAs is confirmed by ac magnetic
susceptibility measurements. Figure 2 (c) shows the the real part
$\chi'$ of ac susceptibility in a temperature range near Tc for
the x=0.16 sample. Figure 3 (a) is the phase diagram showing the
resistivity anomaly (circle) and superconducting transition
(square) temperatures as a function of F content.

\begin{figure}[t]
\centerline{\includegraphics[width=3.2in]{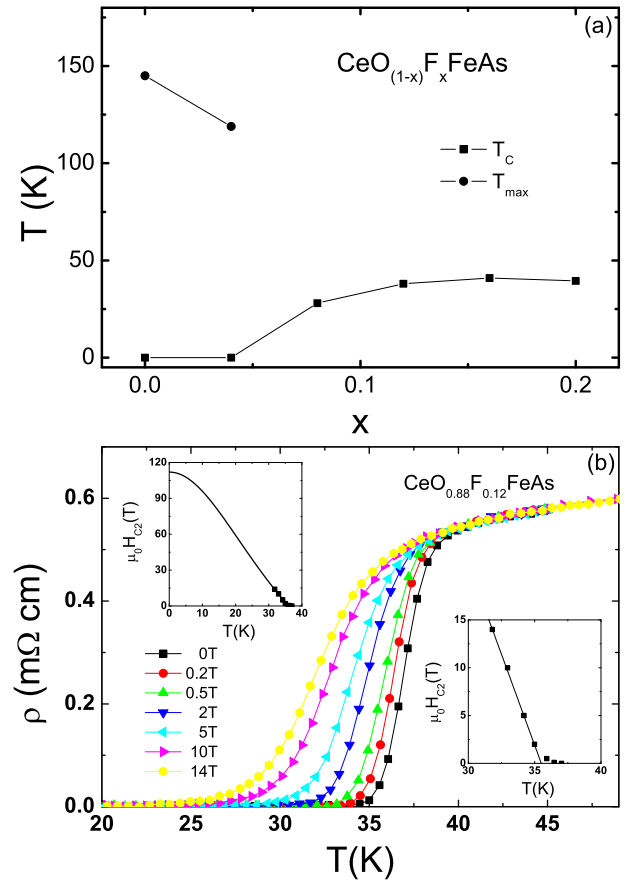}}%
\caption{(Color online) (a) The phase diagram showing the anomaly
(circle) and superconducting transition (square) temperatures as a
function of F content. (b) The resistivity vs temperature curves
under several selected magnetic fields. The lower inset shows the
temperature dependence of upper critical field. The solid line
indicates the slope. The upper inset is a fit to the equation
described in the text.}
\end{figure}

An important parameter to characterize superconductivity is the
upper critical field H$_{c2}(0)$. In our earlier study on
LaO$_{0.9}$F$_{0.1-\delta}$FeAs superconductor with an onset
T$_c$=26 K, we already found a rather high upper critical field
H$_{c2}(0)$$\sim$54 T \cite{Chen}. Here we would expect much
higher H$_{c2}(0)$ in Ce-based compounds owing to their
substantially higher T$_c$. For this purpose, we measured the
temperature-dependent resistivity of a x=0.12 sample with T$_c$
onset close to 40 K under different magnetic fields. As shown in
Fig. 3 (b), T$_c$ was suppressed only by several Kelvins at 14 T
(which is the highest magnetic field available in our PPMS
system). The critical field vs. temperature (H$_{c2}$-T) curve
near T$_c$ was plotted in the lower inset. Here the
T$_c$(H$_{c2}$) is defined as a temperature at which the
resistivity falls to half of the normal state value (middle
transition). As described in our earlier work\cite{Chen}, the
H$_{c2}(0)$ could be extracted from two different methods: (i) by
using the Werthamer-Helfand-Hohenberg (WHH) relation,
H$_{c2}$(0)$\approx$ 0.691$\mid$dH$_{c2}$/dT$\mid$$\times$T$_{c}$,
with a critical-field slope near T$_c$ (being about -3.86 T/K),
(ii) from a fit to the equation
$H_{C2}$(T)=$H_{C2}$(0)[1-(T/T$_{c})^2$]/[1+(T/T$_{c})^2$], as
shown in the upper inset. Then, we obtain H$_{c2}(0)$$\sim$107 T
and 112 T, respectively. We remark here that those rather high
estimated values of H$_{c2}(0)$ are still at the lower limit for
the upper critical field, because the criteria for the
T$_c$(H$_{c2}$) in above analysi is defined at the middle
transition, not at the onset transition temperature, which
actually shows smaller shift with field. Additionally, the
multiple bands effect was not taken into account.

\begin{figure}[t]
\centerline{\includegraphics[width=3.0in]{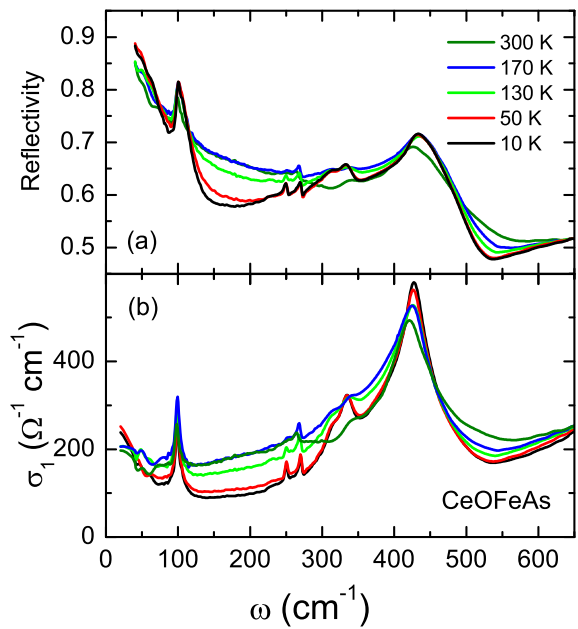}}%
\caption{(Color online) The reflectance (a) and conductivity (b)
spectra in the far-infrared region at different temperatures for
the pure CeOFeAs sample.}
\end{figure}

The resistivity behavior of the pure CeOFeAs is very similar to
that of LaOFeAs, except for the difference that a resistivity
upturn was observed in the later compound at lower temperature. As
we demonstrated earlier, the anomaly at 150 K is caused by
spin-density-wave instability, and a gap opens below the
transition temperature due to the Fermi surface
nesting\cite{Dong}. To confirm the same origin for the anomaly, we
performed infrared measurement on Bruker 66v/s spectrometer in the
frequency range from 40 $cm^{-1}$ to 15,000 $cm^{-1}$ at different
temperatures, and derived conductivity from Kramers-Kronig
transformations. Figure 4 shows the reflectance and conductivity
spectra in far-infrared region. As expected, CeOFeAs shares very
similar optical response behavior as LaOFeAs. Most notably, the
reflectance below 400 cm$^{-1}$ is strongly suppressed at low
frequency below the phase transition temperature, which is a
strong indication for the formation of an energy gap. However, the
low-frequency reflectance still increases fast towards unity at
zero frequency, indicating metallic behavior even below the phase
transition, being consistent with the dc resistivity measurement
which reveals an enhanced conductivity. The data indicate clearly
that only partial Fermi surfaces are gapped.

We noticed that, among different reported superconducting systems
in such layered transition metal oxypnictides, the
LaO$_{1-x}$F$_x$FeAs and CeO$_{1-x}$F$_x$FeAs systems share
remarkably similar phenomenon: the presence of competing ground
states. When the SDW order is destroyed by electron doping,
superconductivity could occur at a much higher temperature. This
gives a hint where to search for materials with potentially higher
T$_c$. The interplay between superconductivity and
spin-density-wave instability thus is of central interest in those
systems.

\begin{figure}[t]
\centerline{\includegraphics[width=3.0in]{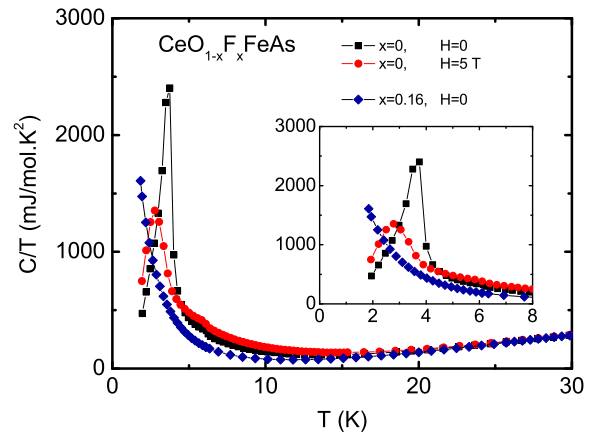}}%
\caption{(Color online) The plot of C/T vs. T for the pure CeOFeAs
sample at H=0 and 5T, and the 16$\%$ F-doped
CeO$_{0.84}$F$_{0.16}$FeAs sample at H=0, respectively. The inset
shows the plot in an expanded low temperature range.}
\end{figure}

To get insight into whether the rare-earth element Ce 4f electrons
hybridize with the itinerant Fe 3d electrons at low temperature,
we measured the low-T specific heat. To our surprise, another
magnetic ordering was revealed in those Ce-based samples. Fig. 5
shows the plot of C/T as a function of temperature for pure
CeOFeAs at H=0 and 5T, and 16$\%$ F-doped
CeO$_{0.84}$F$_{0.16}$FeAs at H=0, respectively. For
non-superconducting CeOFeAs, a sharp $\lambda $-shape peak at 3.7
K is observed under zero magnetic field. The peak shifts to 2.8 K
under a magnetic field of 5 T. This clearly indicates that an
long-range antiferromagnetic ordering transition occurs at low
temperature. There is a very weak effect in \textit{dc}
resistivity at the AFM transition temperature. For LaOFeAs without
rare earth 4f electrons, there is no such specific heat anomaly at
low temperature\cite{Dong}, indicating unambiguously that the AFM
transition for CeOFeAs is originated from the ordering of Ce 4f
moments. No significant enhancement of electronic specific
coefficient is observed from the high T specific heat. For the
16$\%$ F doped superconducting sample, we also observe the onset
signature of the AFM transition down to 1.8K, the lowest measured
temperature. The AFM transition must occur below this temperature.
The data strongly suggest that the high-temperature
superconductivity coexists with the AFM ordering of Ce 4f local
moments. This coexistence implies that the exchange interaction
between the Ce 4f moments and the itinerant Fe 3d electrons is
very weak. So, there is no appreciable mixing or hybridization
between them in the present systems.

To summarize, we have synthesized a series of rear-earth based
transition metal oxypnictide CeO$_{1-x}$F$_x$FeAs compounds. The
superconducting transition temperature could be as high as 41 K.
This very high superconducting transition temperature strongly
challenges the pairing mechanism based on the electron-phonon
interaction. Similar to the LaOFeAs, the pure CeOFeAs shows a
strong resistivity anomaly near 145 K, which was ascribed to the
spin-density wave instability. F-doping suppresses this
instability and leads to the superconducting ground state with
rather high T$_c$. The very interesting interplay between the
superconducting phase and the spin-density-wave instability
strongly suggests that the magnetic fluctuations play a key role
in the superconducting paring mechanism. Furthermore, the study
reveals that the Ce 4f electrons form local moments and ordered
antiferromagnetically below 4 K which could coexist with
superconductivity.

\begin{acknowledgments}
We acknowledge the support from Y. P. Wang and Li Lu, and valuable
discussions with Z. Fang and T. Xiang. This work is supported by
the National Science Foundation of China, the Knowledge Innovation
Project of the Chinese Academy of Sciences, and the 973 project of
the Ministry of Science and Technology of China.

Note added: After we completed this work, we learnt an independent
work on another rare-earth Sm-based SmFeAsO$_{1-x}$F$_x$ (x=0.1)
by Chen et al., arXiv:0803.3603v1.

\end{acknowledgments}


\end{document}